\documentclass[twocolumn,preprintnumbers,amsmath,amssymb,aps,prl,superscriptaddress]{revtex4}
 
\usepackage{graphicx}
\usepackage[caption=false]{subfig}
\usepackage{amssymb}
\usepackage{enumerate}
\begin{document}

\title[Transport through quantum rings]{Transport 
through quantum rings}

\author{B. A. Z. Ant\'{o}nio}
\affiliation{Department of Physics, I3N, University of Aveiro,\\
Campus de Santiago, Portugal}

\author{A. A. Lopes}
\affiliation{Institute of Physics, University of Freiburg, Hermann-Herder-Straße 3, 79104 Freiburg, Germany.}

\author{R. G. Dias}
\affiliation{Department of Physics, I3N, University of Aveiro,\\
Campus de Santiago, Portugal}

\begin{abstract}
The transport of fermions through nanocircuits plays a major role in mesoscopic physics. 
Exploring the  analogy with classical wave scattering,
basic notions of nanoscale 
transport can be explained in a simple way, even at the level of undergraduate Solid State Physics courses, and more so if these explanations are supported by numerical simulations of these nanocircuits. 
This paper presents a simple tight-binding method for the study of the conductance of  quantum nanorings connected to one-dimensional leads. We show how to address the  effects of applied magnetic and electric fields and illustrate concepts such as Aharonov-Bohm conductance oscillations, resonant tunneling and destructive interference.
\end{abstract}

\maketitle

\section{Introduction}
Constructing  circuits using  nanoscale components  such as single electron 
transistors, single molecule devices, nanowires etc, seems to be the most 
plausible  way of extending Moore's Law for a few more years.
Electron transport in these  circuits must be addressed in the quantum 
mechanical framework and surprising features (strongly  contrasting  with 
those of  bulk electronic transport) are  known to occur, e.g. Coulomb blockade \cite{Gorter1951}, 
conductance quantization \cite{Wees1988} and resonant tunneling \cite{Stone1985}. Much of the underlying physics 
behind these phenomena can be explained in a simple way, at least qualitatively, 
by analogy with classical wave scattering and  
it is reasonable to expect a growing attention to these basic notions of nanoscale 
transport in the curricula of undergraduate Solid State Physics courses.

One of the simplest and most interesting nanocircuits is the open ring, 
for its simplicity and non-local quantum effects. 
Particle waves incident in the ring travel by both arms of the ring and 
interfere constructively or destructively on their way out, depending on 
the magnetic flux encircled by the quantum ring and on the position of the contacts. This leads to the 
so-called Aharonov-Bohm oscillations in the conductance \cite{Aharonov1959,Webb1985}.
In this paper we present a simple method for the numerical calculation  
of the conductance through tight-binding 
clusters and apply it to the particular case of quantum rings taking into 
account the dependence on applied magnetic and electric 
fields. In particular, we discuss the Aharonov-Bohm  conductance oscillations. 
The leads are assumed to be one-dimensional for simplicity.
We avoid defining quantities such as  Green's functions or self-energies and  
the problem of determining the conductance    is reduced to that of finding the solution of    a system of $N+2$ linear equations (where $N$ is the number of sites of the ring), so that this computational study 
can be carried out by an  undergraduate student, after taking an  
introductory course in quantum mechanics.

The remaining part of this paper is organized in the following way. 
In Sec.~\ref{section:system}, we describe the Hamiltonian for the system of the 
cluster and leads. In Sec.~\ref{section:conductance}, we show how to reduce the determination of the conductance to that of  finding the solution of a linear system of $N+2$ equations. In Sec.~\ref{section:results}, we present the 
results for the conductance through tight-binding  rings in the presence of  
magnetic flux and electric fields. Finally in Sec.~\ref{section:conclusion} we 
draw the conclusions of this work.
\section{System}
\label{section:system}

Our system is composed by a tight-binding cluster with $N$ sites connected to 2 semi-infinite 
1D leads modelled as 1D tight-binding semi-chains  \cite{Marder2010}. The system Hamiltonian is 
the sum of the leads and cluster  Hamiltonians, $H_0$, and the coupling between 
each lead and the cluster, $V_{LR}$, 
\begin{equation}
\label{condu0}
\ H=H_0+V_{LR},
\end{equation}
with 
\begin{equation}
    \begin{split}
        \label{condu1}
        H_0 &= \displaystyle
       \underbrace{ 
       -\sum\limits_{j=-\infty}^0  
       \left(\vert j-1\rangle \langle j\vert +\mathrm{H.c.} \right)}_{\mathrm{left \,lead}}
       - \underbrace{\sum\limits_{j=N+1}^\infty \left(\vert j\rangle \langle j+1\vert 
       +\mathrm{H.c.}  \right)}_{\mathrm{right\, lead}} \\
       &+\underbrace{H_S}_{\mathrm{cluster}} 
   \end{split}
\end{equation}
where $H_s$ is the single-particle Hamiltonian in the scattering region 
(the cluster) and $\vert j \rangle$ is the  Wannier state localized on the atomic site $j$ (see Ref.~\cite{Marder2010} for an introduction to tight-binding models using the Wannier function basis). The leads connected to the cluster are considered ideal 
with nearest-neighbour hopping $t=1$. 
The coupling $V_{LR}$ between the leads and the cluster is given by
\begin{equation}
    \label{condu2}
    V_{LR}=-t_L\vert 0\rangle\langle L\vert-t_R\vert R\rangle\langle N+1\vert 
    +\mathrm{H.c.} ,
\end{equation}
where the  hopping matrix elements $t_L$ and $t_R$ connect the left and right leads, 
respectively, to the cluster sites $L$ and $R$. 
\begin{figure}[t]
	\centering
	\includegraphics[width=0.4\textwidth]{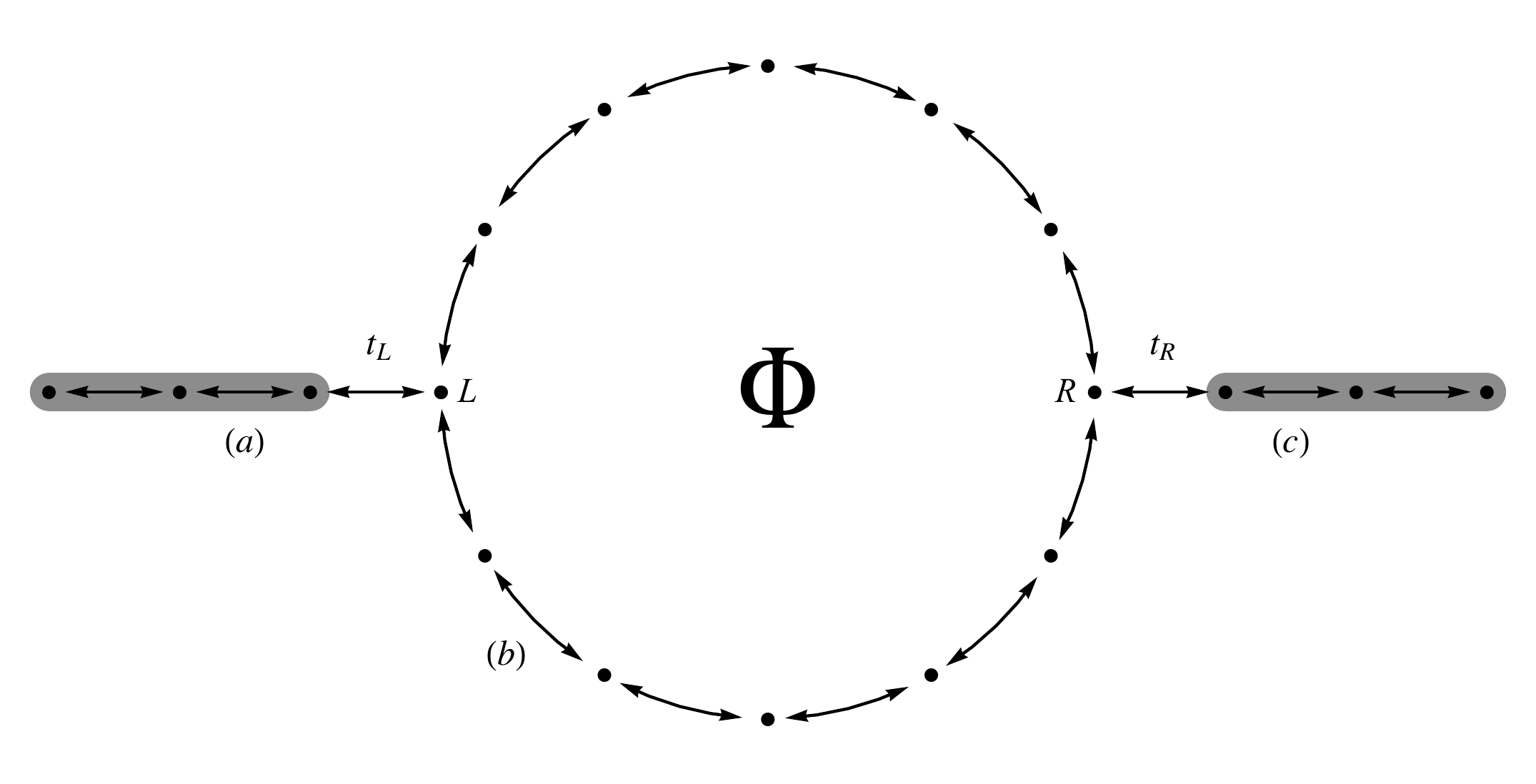} 
	\caption{Schematic representation of the studied systems. In (a) and (c) 
		one has the left and right semi-infinite 1D leads, respectively. 
		These leads are considered to be perfect conductors. In (b), we 
		show the cluster described by a tight-binding model.  The hopping factors which couple 
		the left and right leads to the cluster are, respectively, $t_L$ and $t_R $.}
	\label{systemschematic}
\end{figure}
The Hamiltonian of the cluster is 
\begin{equation}
    \label{AB5}
     H_S= \displaystyle\sum\limits_{j}
     {\epsilon}_j \vert j\rangle \langle j\vert - 
     \sum\limits_{ij, i\neq j} \left( t_{ij}e^{i\phi_{ij}}
     \vert i\rangle \langle j\vert +\mathrm{H.c.}\right)
\end{equation}
where the indices $i$ and $j$ run over all cluster sites.
For a general gauge and an arbitrary geometry of the cluster, 
the phase shifts $\phi_{ij}$ in the hopping factors between sites 
$\mathbf{r}_i=\mathbf{a}$ and $\mathbf{r}_j=\mathbf{b}$ are determined from
$
\phi_{ij}=-\frac{e}{\hbar}\int_{\mathbf{a}}^{\mathbf{b}}{\mathbf{A}. d\mathbf{r}}
$
where the integral is a standard line integral of the vector potential 
$\mathbf{A}$ along the line segment  which links the sites a and b.

\begin{figure}[t]
	\subfloat[]{\label{fig:perfect1} 
	\includegraphics[width=.25\textwidth]{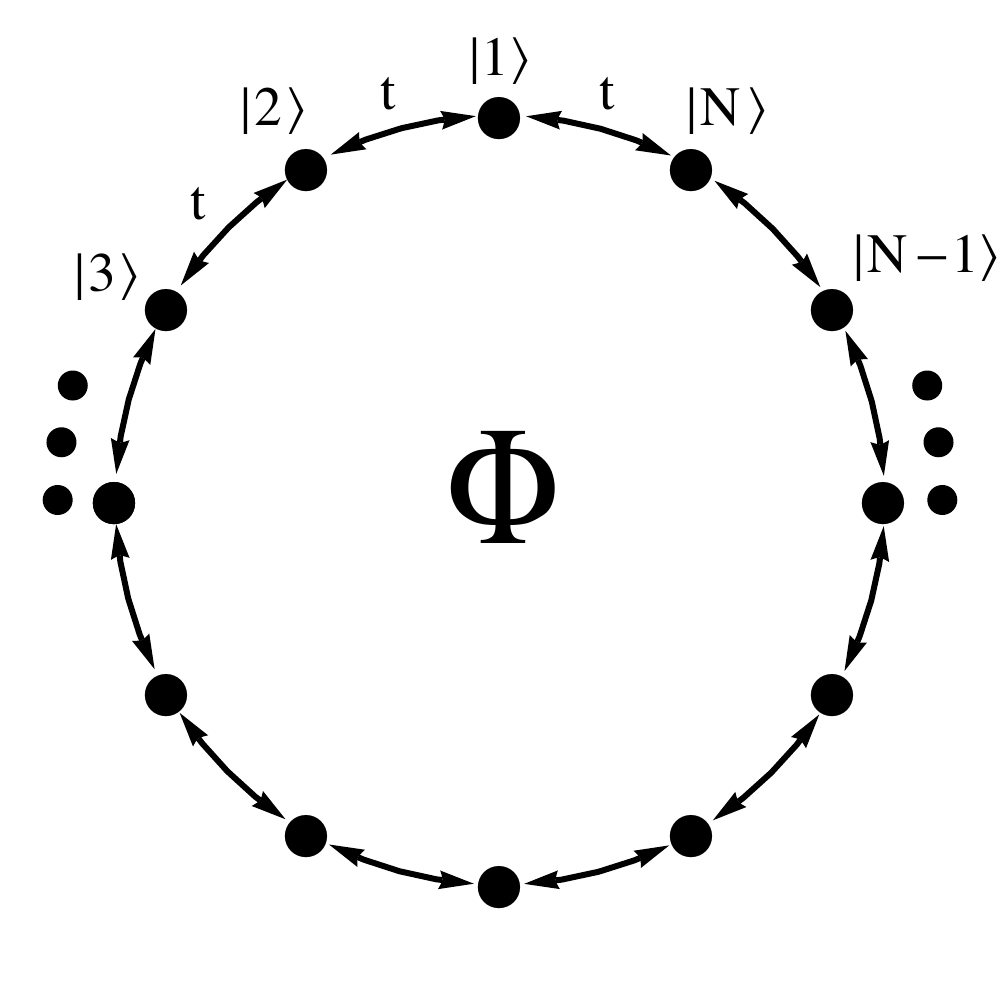} }\\
	\subfloat[]{\label{fig:theorethicalspectra} 
	\includegraphics[width=.3\textwidth]{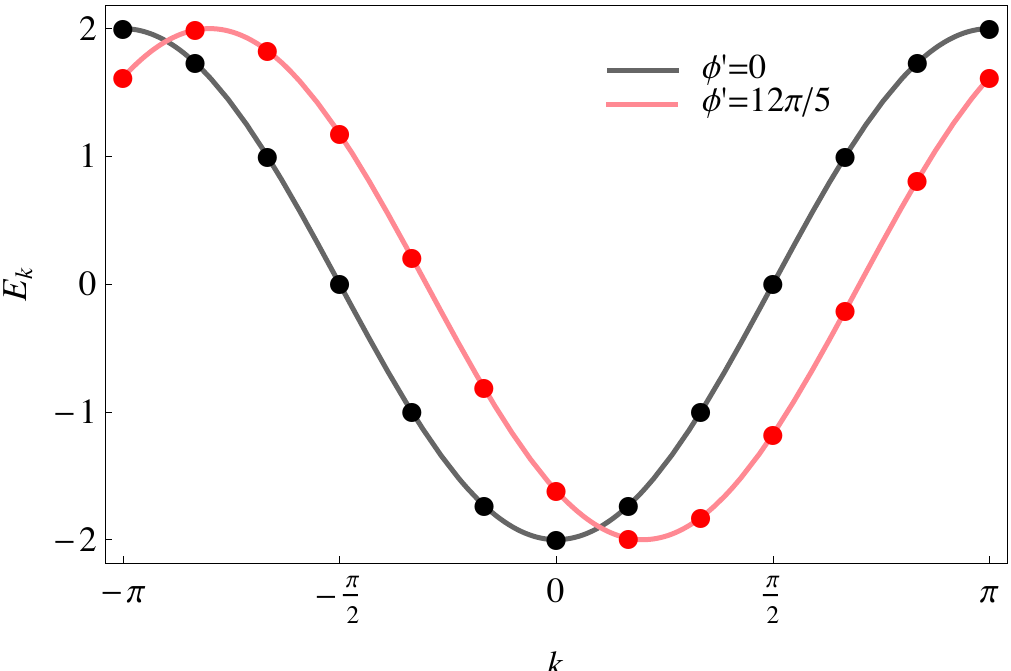} } 
	\caption{(a) Schematic diagram of a perfect ring with $N$ sites, 
		where the sites are numbered in an anti-clockwise way, 
		threaded by magnetic flux $\phi$. (b) Energy spectra of the ring 
		with 12 sites with and without magnetic flux. 
		The flux shifts the energy dispersion curve.}
	\label{fig:fluxos}
\end{figure}

In this paper, we exemplify the application of the  numerical method by using  the particular case of  the conductance through a 
tight-binding ring threaded by a magnetic flux. This is a well studied case and the reader should see for example  Refs.~\cite{Li1997,Kowal1990,Orellana2003}  for  in-depth  analyses of the tight-binding ring  conductance.
The tight-binding Hamiltonian 
of a ring of $N$ sites enclosing an external magnetic flux  can be expressed as
\begin{equation}
    \label{AB5.1}
     H_S= \displaystyle\sum\limits_{j=1}^N 
     {\epsilon}_j \vert j\rangle \langle j\vert - 
     \sum\limits_{j=1}^N t\left(e^{i\phi ' /N}
     \vert j\rangle \langle j+1\vert +e^{-i\phi ' /N}
     \vert j+1 \rangle \langle j\vert \right)
\end{equation}
where $\phi'=2\pi\Phi/\Phi_0 $ is the reduced flux, 
$\epsilon_j$ is the on-site energy,
$t$ is the hopping integral  between two nearest neighbour sites, and 
$\Phi_0=h/e$ is the flux quantum  \cite{Marder2010}.
A particular  gauge was chosen such that the  Peierls phases are all the same 
and, assuming that the on-site  energies $\epsilon_j$ are site independent,
the tight-binding Hamiltonian becomes  translationally invariant  and therefore 
its eigenfunctions are given by
\begin{equation}
    \label{tight4}
    \vert k \rangle = \frac{1}{\sqrt{N}} 
    \displaystyle\sum\limits_{j=1}^N e^{ikj} \vert j\rangle
\end{equation}
where $\{\vert k\rangle\}$ is the momentum basis, 
$N$ is the number of atoms in the ring, $k$ is the particle momentum and 
$\{\vert j\rangle\}$ is the position basis of the atoms.
The periodic boundary condition leads to the momentum's quantization,
$
    k=2\pi m/N,\, m=0,1,2,N-1,
$
and the eigenvalues of the Hamiltonian are given by 
\begin{equation}
    \label{AB6}
    E(k)=-2t\cos \left(k-\frac{\phi'}{N}\right).
\end{equation}
where we consider the on-site energies to be equal  to zero.
The flux shifts the energy dispersion curve, as shown in Fig.~\ref{fig:fluxos}.
 
\section{Quantum scattering, transmittance and conductance}
\label{section:conductance}
In this section, we explain how to obtain the conductance expression for  a tight-binding 
cluster connected to two semi-infinite (modelled as tight-binding chains) 
leads.  Our approach requires only the understanding of the tight-binding model which is usually introduced in an undergraduate Solid State course (see Refs.~\cite{Marder2010,datta2005} for a  discussion of tight-binding models).
We avoid therefore the use of Green's functions which are present in more  advanced conductance calculations such as  the Meir-Wingreen formulation \cite{Meir1992} (which is based on the non-equilibrium Green's functions method and  takes into account electronic interactions in the cluster)
or the traditional quantum scattering approach \cite{Taylor1972, Enss2005}, (which is a one-particle description of scattering events due to potential barriers).
Our approach is further  simplified by the fact that the particles are confined to 1D leads, and consequently no angle dependence is present in the conductance expressions.

In nanoscale devices   the wavelike nature of the electron must be taken 
into account. In this case the quantum scattering description of  the transmission 
of a 1D particle across a structural or potential barrier  is closely analogous 
to the Fresnel description of a plane light wave incident in a interface between 
two dielectric media. 
The wavefunction of a particle with momentum $k$ and energy $\epsilon_k$ is 
written as an incident wave plus reflected and transmitted waves with 
amplitudes $r(\epsilon_k)$ and $t(\epsilon_k)$ respectively, 
relatively to the incident wave.
The transmission probability for this incident wave is $\vert t(\epsilon_k) \vert^2$.
The linear conductance $G(T)$ of non-interacting fermions is determined using the 
Landauer formula \cite{Landauer1970}, which relates the transmission 
probability $|t(\epsilon)|^2 $ with the linear conductance $G(T)$ of the 
non-interacting fermions at temperature $T$. 
This formula is explained in the following way.
A single state with a momentum $k$ (and an energy  
$\epsilon_k=-2 \cos k$ of the band of a 1D lead) carries a current 
$ev_k/L $ where $v_k$ is the 1D velocity of the electron, 
$\hbar v_k=\partial \epsilon_k /\partial k$ and $L$ is the length of the lead.
The contribution to the current of the positive velocity states 
in an interval $\Delta k$ is therefore 
$I_{\Delta k} =e/(\hbar L) \sum_{k \, \Delta k} d \epsilon_k /d k$. 
Assuming $L$ very large, so that the $k$ values are effectively continuous, 
the summation over $k$ can be replaced by an integral 
$\sum_k \rightarrow (L/2 \pi) \int dk$ and converted into an integral 
over energy  $ \int dk = \int (dk/d \epsilon_k) d\epsilon_k$. In the final 
expression for the current, the two derivatives cancel each other  and one has
$I_{\Delta \epsilon}= (e/h) \Delta \epsilon$ where $\Delta \epsilon$ 
is the energy interval corresponding to $\Delta k$. This result states 
that independently of the form of the 1D band, all energy intervals with 
the same width  within the bandwidth contribute equally to the current. 
In the case of the 1D tight-binding chain, one has that the velocity of the 
electrons goes to zero as  the energy approaches the top or the 
bottom of the band and therefore the contribution of an individual 
state to the current goes also to zero. This is however compensated 
by the fact that the density of states is inversely proportional to the 
velocity and therefore diverges as the energy approaches the top or 
the bottom of the band, compensating the smaller individual contribution of each state.
The same reasoning is followed for negative velocity states which 
generate a current in the opposite direction. A finite bias implies 
that the chemical potential for electrons with positive or negative 
velocities is different (they originate from opposite ends of the system), 
$\mu_1$ and $\mu_2$ respectively, and the zero temperature net 
current is given by $I= (e/h) (\mu_1- \mu_2)$. The respective conductance is $G=I/V=I/[(\mu_1- \mu_2)/e]=e^2/h$.
At finite temperature the bands are populated according to the Fermi-Dirac 
distribution $f(\epsilon,\mu)=[\exp[(\epsilon-\mu)/k_BT]+1]^{-1}=f_0(\epsilon - \mu)$ and the current 
becomes $I= (e/h) \int_{-\infty}^\infty [f_0(\epsilon-\mu_1)- f_0(\epsilon-\mu_2))$.
The final ingredient is the introduction of an obstacle (a structural or 
potential barrier) in the path of the 1D electrons. An electron with energy $\epsilon$  is transmitted through the barrier 
with probability $|t(\epsilon)|^2 $ as explained above and the current becomes 
$I= (e/h) \int_{-\infty}^\infty |t(\epsilon)|^2 [f_0(\epsilon-\mu_1)- f_0(\epsilon-\mu_2))$.
Since  $\mu_1 -\mu_2=eV$, the differential conductance $G=dI/dV$ 
for spinless fermions on the lattice is given by
\begin{equation}
    \label{condu3}
    G(T)=\displaystyle\frac{e^2}{h}\int\limits_{-2}^{2}
    \left(-\frac{df}{d\epsilon}\right)|t(\epsilon)|^2 d\epsilon ,
\end{equation}
where  $f=f(\epsilon,\mu)$ is  the 
Fermi-Dirac distribution \cite{Landauer1970}.
For zero temperature, 
$
    G(0)=\frac{e^2}{h} |t(\mu)|^2.
$
The energy interval  of integration corresponds to  the energies of the 
one-particle eigenstates of the 1D tight-binding leads with hopping integral $t=1$, 
since  for time much earlier or later than when the incident  wave packet  
is in the neighbourhood of the cluster,  the particle can be 
described as a 1D tight-binding particle. That is, the energy spectra  
(and also its density of states) for extended states in the system of  
coupled leads and cluster remains  the same as for the leads without 
coupling to the cluster.

The hoppings $t_L$ and $t_R$ generate  finite transmission probability 
across the cluster. Since no two-particle interactions are considered in this paper, the transmission probability $|t(\epsilon_k)|^2 $ for an incident particle with momentum $k$ and energy $\epsilon_k=-2 \cos (k)$  can be calculated 
solving directly the respective eigenvector equation for the tight-binding Hamiltonian, $ \hat H \vert \Psi_k \rangle =\epsilon_k \vert \Psi_k \rangle$.
This method is  similar to that followed in the determination of the band structure of a tight-binding model, which is usually taught in  a undergraduate Solid State course.
Since $\vert \Psi_k \rangle=\sum_{n=\infty}^\infty \psi_n \vert n\rangle$ where $\psi_n$ is the eigenfunction amplitude at site $n$, the eigenvector equation can be written as a matrix equation
$[H] (\psi) = \epsilon_k (\psi) $, where $(\psi)$ is the column vector of the eigenfunction amplitudes $\psi_n$ and  $[H]$ is the matrix representation of the Hamiltonian in the Wannier function basis $\{ \vert n \rangle \}$. This matrix equation is equivalent to  an infinite system of linear equations (the number of equations is equal to the number of sites which is infinite in our system) of the form
\begin{equation}
\epsilon_k\psi_n = \sum_{m=-\infty}^\infty H_{nm} \psi_m
\end{equation}
where $H_{nm}=\langle n \vert  \hat H \vert m \rangle$. The matrix element $H_{nm}$ is zero except if the site $m$ is a nearest-neighbor of site $n$. 
One  should  now recall that  our system is constituted by the left lead, the cluster and the right lead. If $t_L=t_R=0$, the system of equations decouples into  three independent sets of equations and the particle can be restricted to one of the three regions. For instance, the eigenvectors and eigenvalues when the   particle is confined to the cluster are obtained from $ [H_S ] (\psi_S) =\epsilon (\psi_S)$  where $(\psi_S)$ is the column vector of the eigenfunction amplitudes at the cluster sites and the corresponding equations in the case of the tight-binding ring are of the form
\begin{equation}
\epsilon \psi_j =  {\epsilon}_j \psi_j- 
      t e^{i\phi ' /N} 
     \psi_{j+1} -t e^{-i\phi' /N}
     \psi_{j-1}
     \label{eq:numer1}
\end{equation}
where $j=1, \cdots , N$ and with periodic boundary conditions, $\psi_{N+1}=\psi_1$.

Let us now assume  finite $t_L$ and $t_R$.
Since we are interested in the transmission probability of a right-moving particle, we can limit our study to states with energy $\epsilon_k=-2 \cos (k)$ which can be written as an incident wave    plus the respective reflected wave on the left lead and a transmitted wave on the right lead.
This implies that the equations for the wavefunction amplitude at any site $j$ of the leads (with $j<0$ or $j>N+1$),
\begin{equation}
\epsilon_k \psi_j =  -  \psi_{j+1} - \psi_{j-1},
\end{equation}
are automatically satisfied if the wavefunction in the leads is of the form 
\begin{eqnarray}
\psi_j &=e^{ikj}+\psi_r e^{-ikj}, & \quad j\leq 0,
\label{eq:wavereflec} \\
\psi_j &=\psi_t e^{ikj}, & \quad j\geq N+1,
\label{eq:wavetrans}
\end{eqnarray}
where $\psi_r$ and $\psi_t$ are the amplitudes of the reflected and transmitted waves, respectively.
So these equations can be dropped from the the global system of equations. We are left with the equations for the amplitudes at sites $0$ and $N+1$ which involve the hopping constants $t_L$ and $t_R$ and depend on the amplitudes $\psi_L$ and $\psi_R$ at the sites $L$ and $R$ of the ring,
\begin{eqnarray}
\epsilon_k \psi_0 &=&  - t_L \psi_{L} - \psi_{-1},  \label{eq:numer2} \\
\epsilon_k \psi_{N+1} &=&  - \psi_{N+2} - t_R \psi_{R}. \label{eq:numer3}
\end{eqnarray}
The amplitude equations for the cluster  sites remain the same when  $t_L$ and $t_R$ are finite, that is, $ [H_S ] (\psi_S) =\epsilon_k (\psi_S)$, except for the equations corresponding to sites $L$ and $R$ which have an additional term $-t_L \psi_0$ and $-t_R \psi_{N+1}$ and are given by
\begin{eqnarray}
\epsilon_k \psi_L &=&  -t_L \psi_0 +{\epsilon}_L \psi_L- 
      t e^{i\phi ' /N} 
     \psi_{L+1} -t e^{-i\phi' /N}
     \psi_{L-1},  \label{eq:numer4}\\
\epsilon_k \psi_{R} &=&  -t_R \psi_{N+1} +{\epsilon}_R \psi_R- 
      t e^{i\phi ' /N} 
     \psi_{R+1} -t e^{-i\phi' /N}
     \psi_{R-1}.  \label{eq:numer5}
\end{eqnarray}
The solution of this set of $N+2$ equations (that is, Eqs.~\ref{eq:numer1}, \ref{eq:numer2}, \ref{eq:numer3}, \ref{eq:numer4}, and \ref{eq:numer5}) allows us to determine $\psi_t$ and $\psi_r$. Note that $\psi_0$, $\psi_{-1}$, $\psi_{N+1}$ and  $\psi_{N+2}$ are given by Eqs.~\ref{eq:wavereflec} and \ref{eq:wavetrans} and 
are  functions of $\psi_r$ and $\psi_t$, therefore we have $N+2$ variables. 
The transmission probability  is then given by the square of the absolute value of the ratio between the   amplitude of the  outgoing wave $\psi_t$ and the amplitude of the incident wave (which we have assumed to be 1). 
One can easily code  an algorithm  (using for example computational software such as Mathematica or Matlab) that 
allows to: (i) define the  external parameters (magnetic field, electric field, 
temperature and chemical potential); (ii) solve the linear system of equations; (iii)  calculate the transmission probability and the conductance, given as 
input the value of the chemical potential, the cluster contact sites $L$ and $R$  
and the characteristics of the 
cluster:  (a) a list of coordinates $\mathrm{r}_i=(x_i,y_i,z_i)$ of the lattice 
sites $i=1, \cdots, N$ of the cluster; (b) a list of on-site energies 
$\{\epsilon_i\}$; (c) a list of hoppings constants $\{t_{ij}\}$; (d) values 
of the hopping constants $t_L$ and $t_R$. Note that the  $\{\epsilon_i\}$ 
and $\{t_{ij}\}$ lists may include the effect of an external electric field 
or the presence of magnetic flux.

\begin{figure*}[t]
	\begin{center}
	\hspace{-.5cm}	
	\subfloat[]{\label{fig:8ringelectric}
	\includegraphics[width=.3\textwidth]{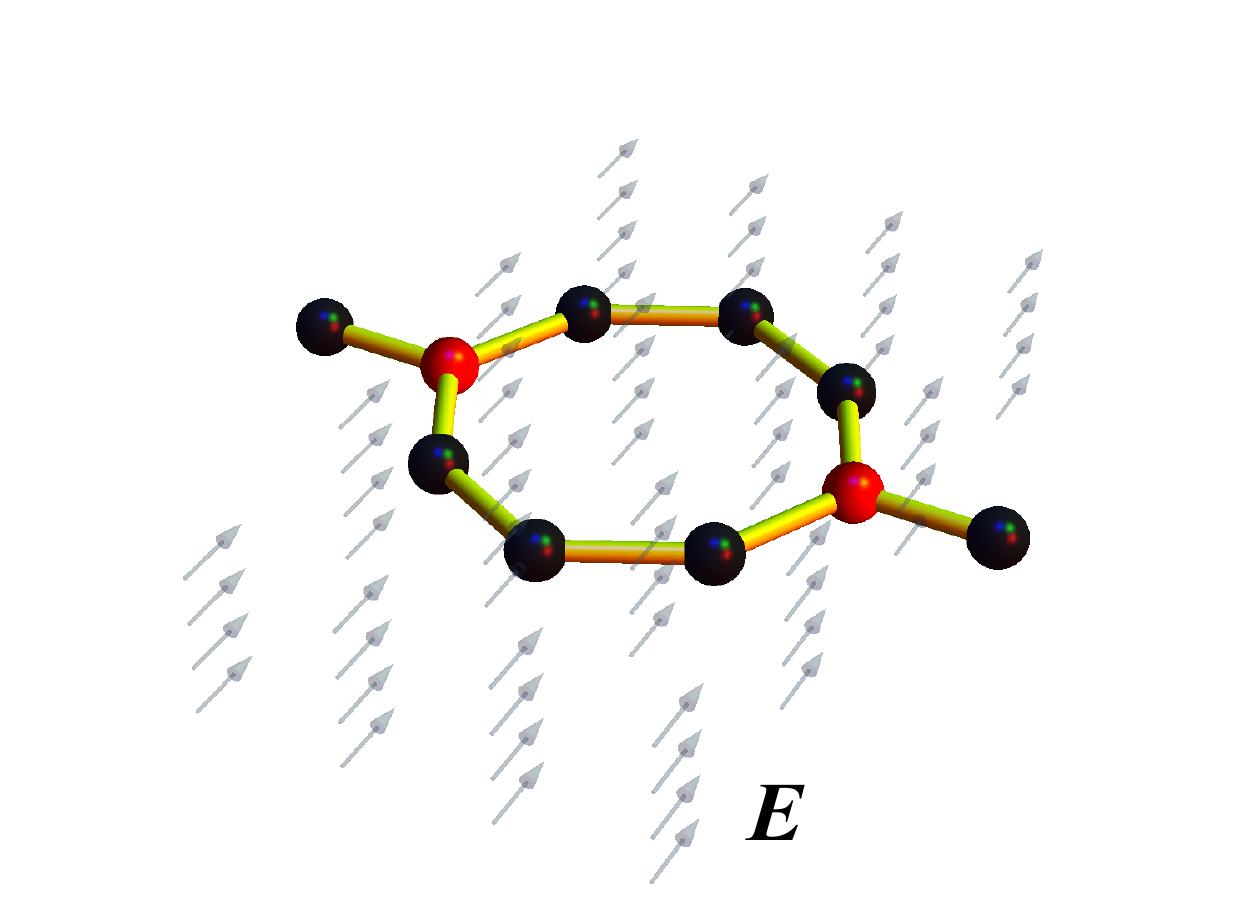}} 
	\hspace{.5cm}
	\subfloat[]{\label{fig:8ringperfect}
	\includegraphics[width=.22\textwidth]{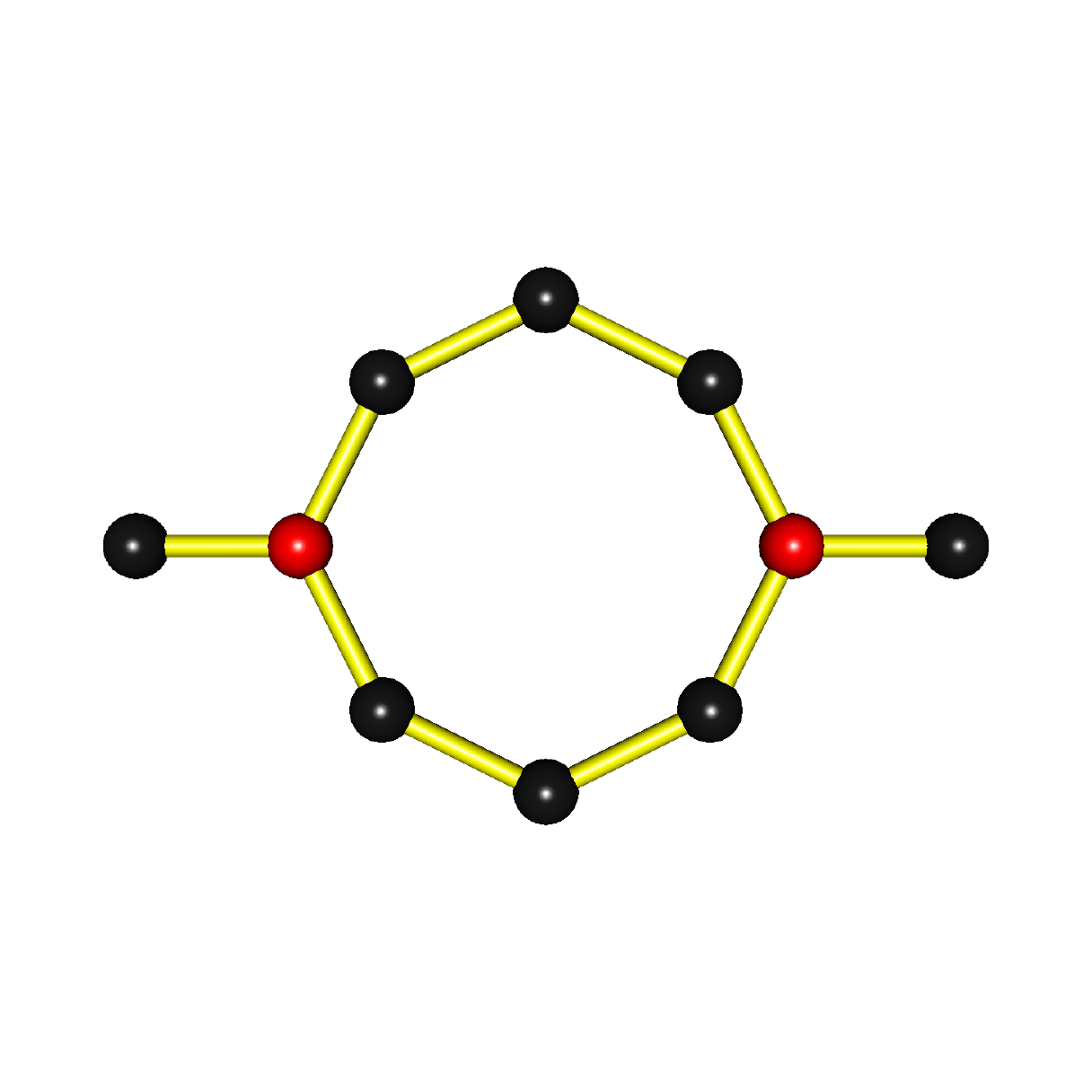}} 
	\hspace{1cm}
	\subfloat[]{\label{fig:8ringincompleto}
	\includegraphics[width=.22\textwidth]{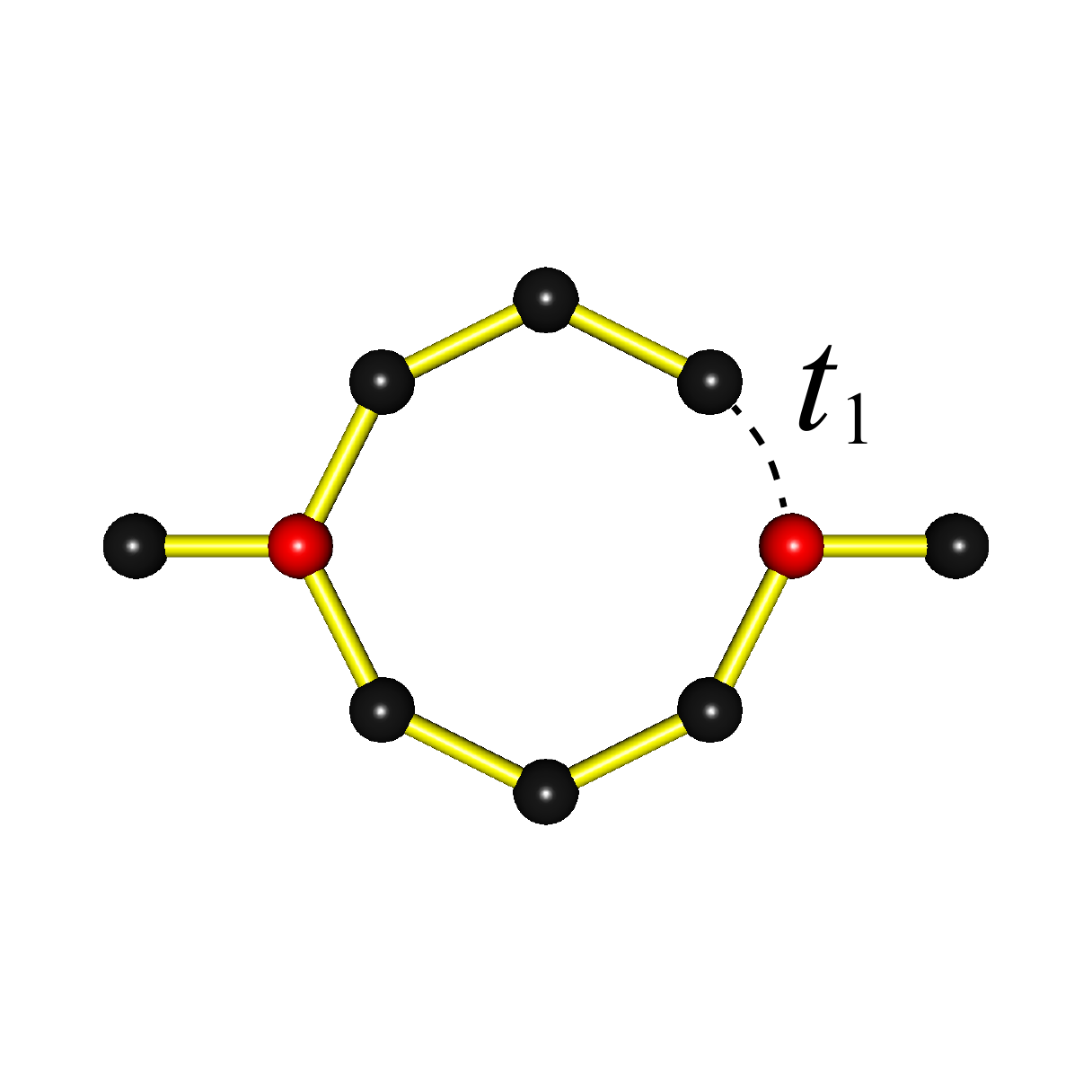} }\\
	\subfloat[]{\label{fig:spectra8ring}
	\includegraphics[width=1\textwidth]{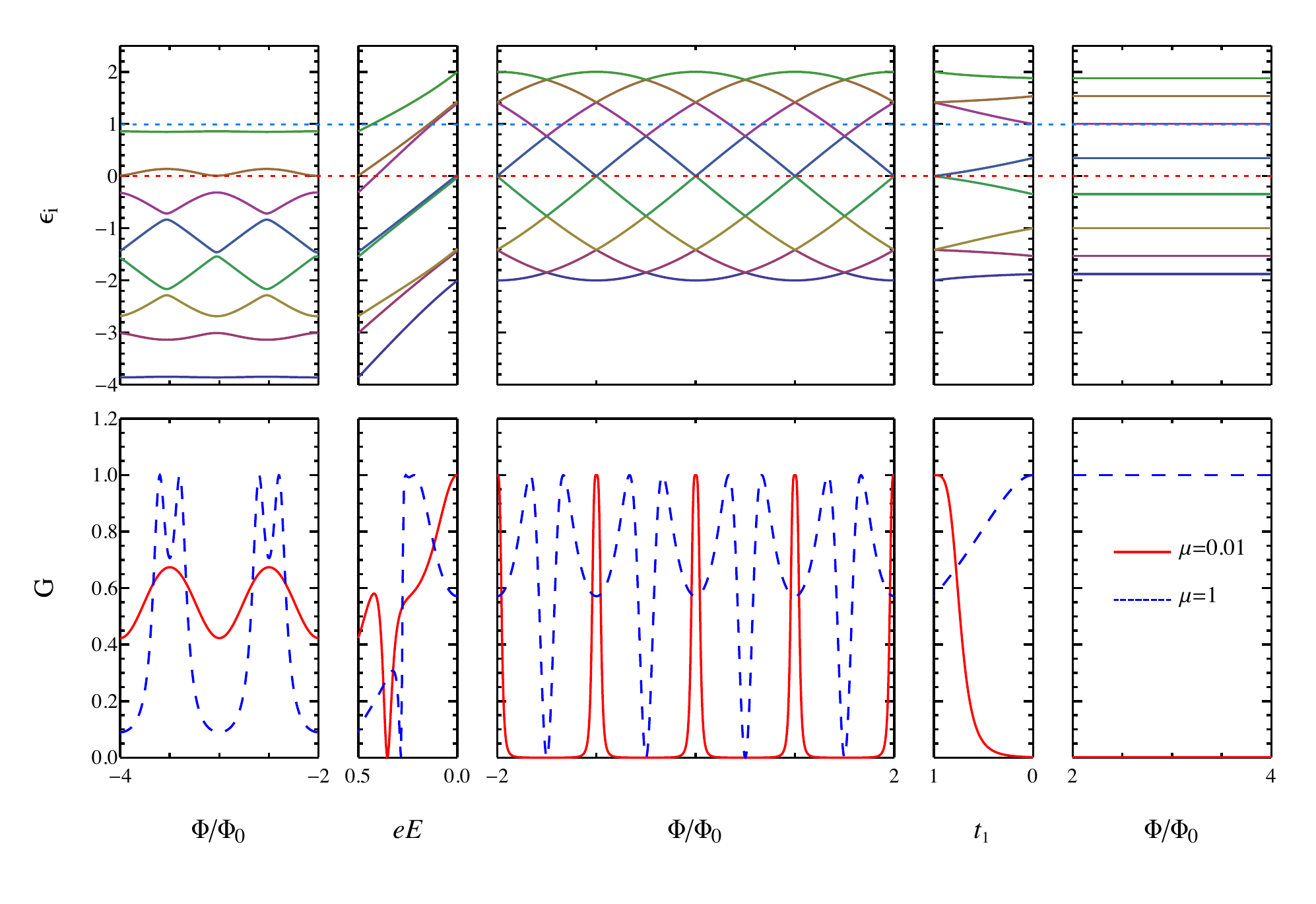} } 
	\caption{
		Schematic representation of a ring with 8 sites: (a) in the presence of an in-plane uniform electric field ($eE_y=0.5$); (b) in the absence of an electric field and (c) with one of the hopping integrals, $t_1$, equal to zero. The contact sites to the leads are indicated in red.
(d) Respective plots of the energy levels and conductance as functions of the normalized magnetic flux $\Phi/\Phi_0$.
The conductance is evaluated for two values of chemical potential, $\mu=0.01$ and 1. Red and blue dashed lines indicate these values of energy in the top plots. 
In the center plot, the Aharonov-Bohm effect in the conductance with a half quantum flux periodicity is observed as well as zero conductance for $\Phi/\Phi_0=n+1/2$ (reflecting the destructive interference of the plane waves travelling through the two branches of the ring). The Aharonov-Bohm effect disappears in the broken ring, as expected.
}
	\label{fig:8aneis}
	\end{center}
\end{figure*}

\begin{figure*}[h]
	\centering
	\subfloat[$\mu=-0.34$,  $E_y=0$]{\label{fig:a} 
	\includegraphics[width=0.3\textwidth]{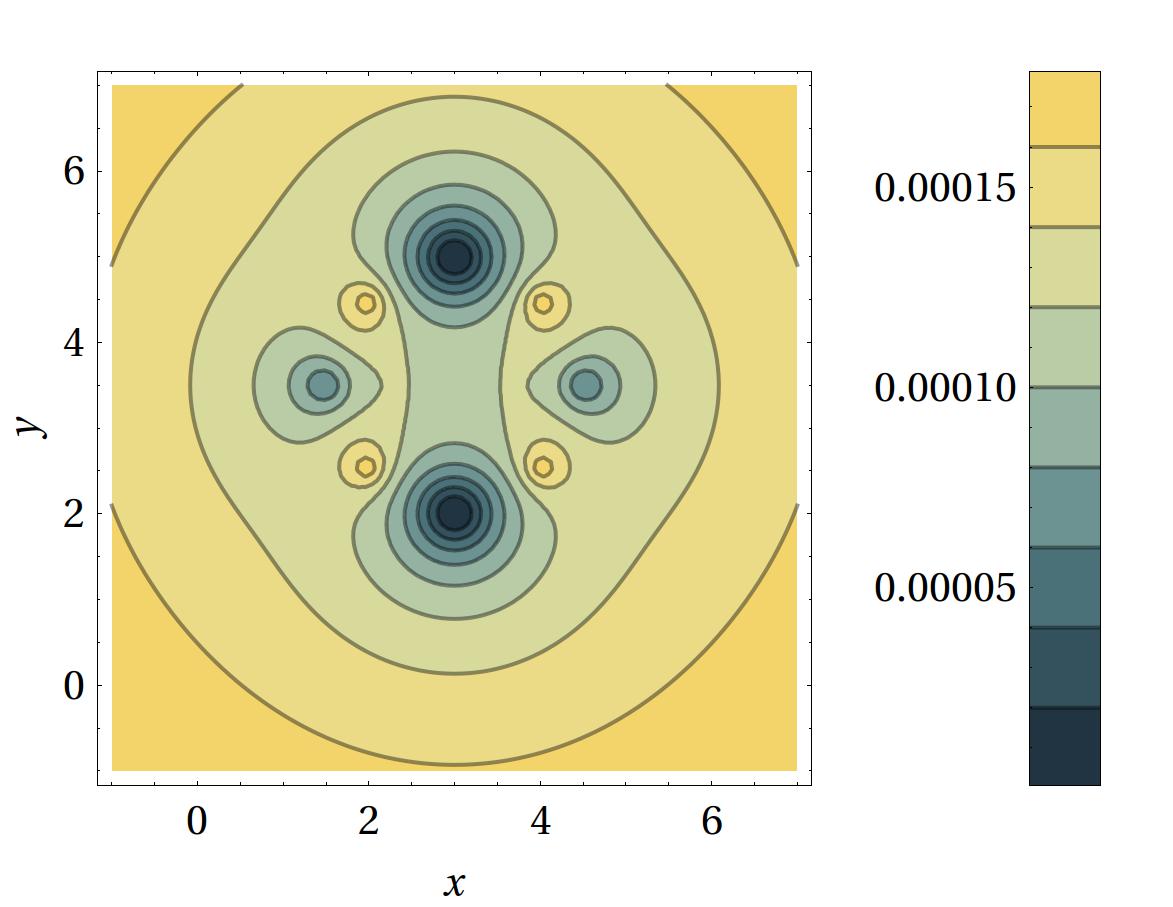} }
	\subfloat[$\mu=0.60$,  $E_y=0$]{\label{fig:b} 
	\includegraphics[width=0.3\textwidth]{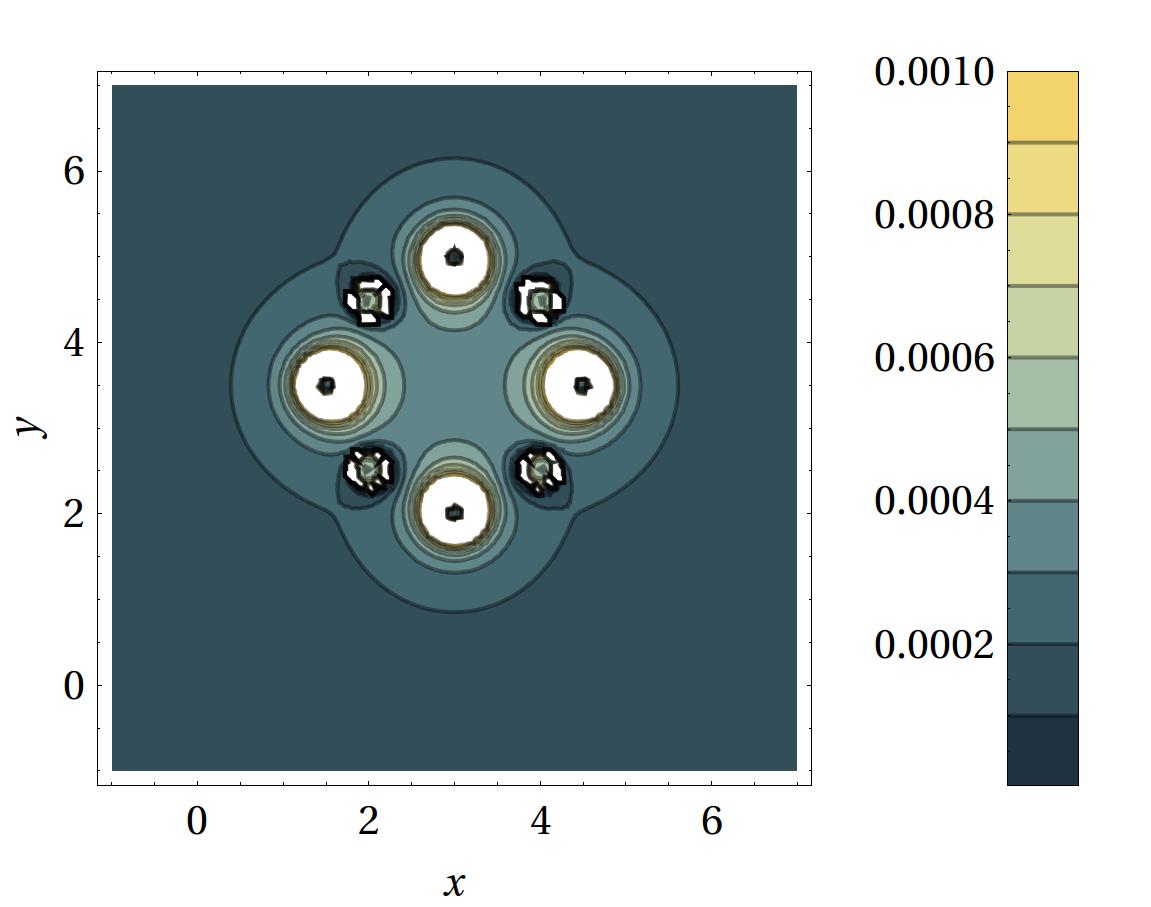} }
	\subfloat[$\mu=-0.76$,  $E_y=0$]{\label{fig:c} 
	\includegraphics[width=0.3\textwidth]{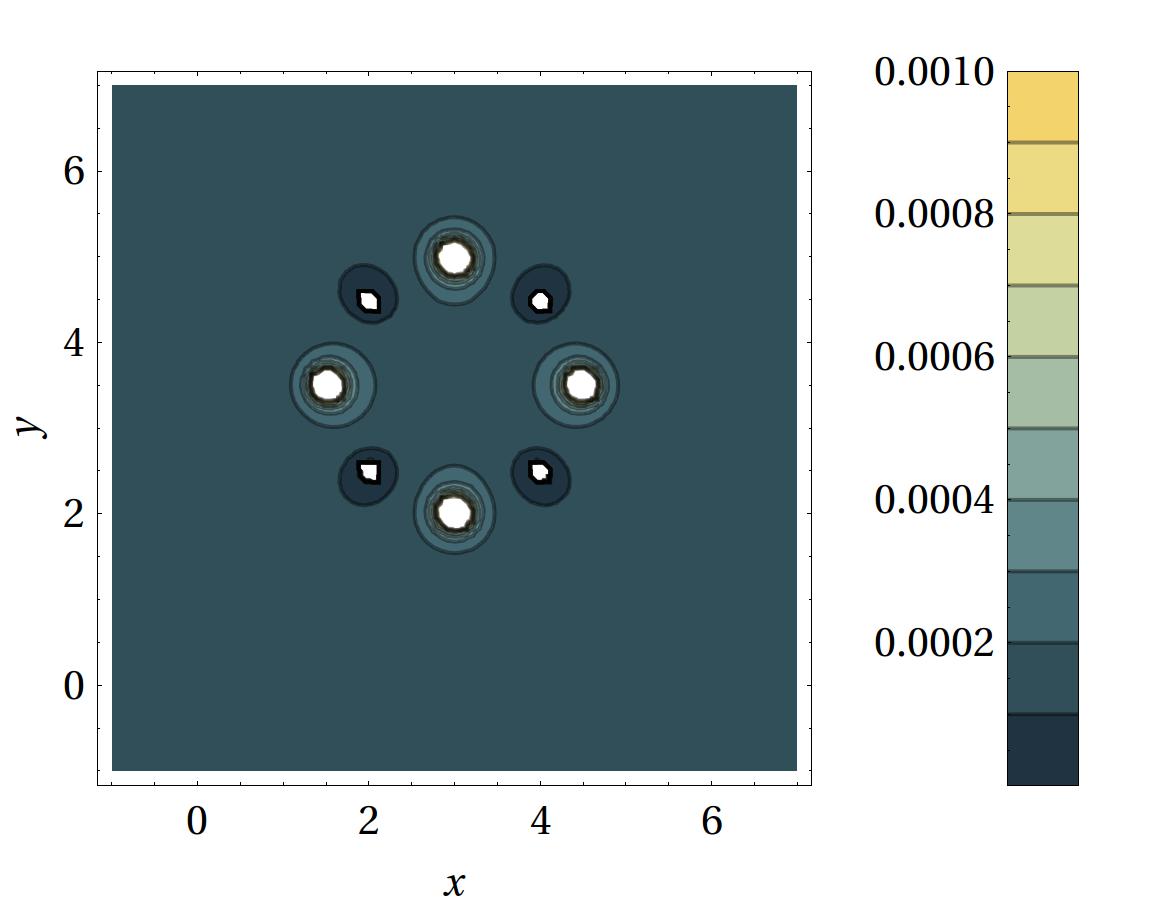} }\\
	\subfloat[$\mu=1$, $E_y=0.46$]{\label{fig:d} 
	\includegraphics[width=0.3\textwidth]{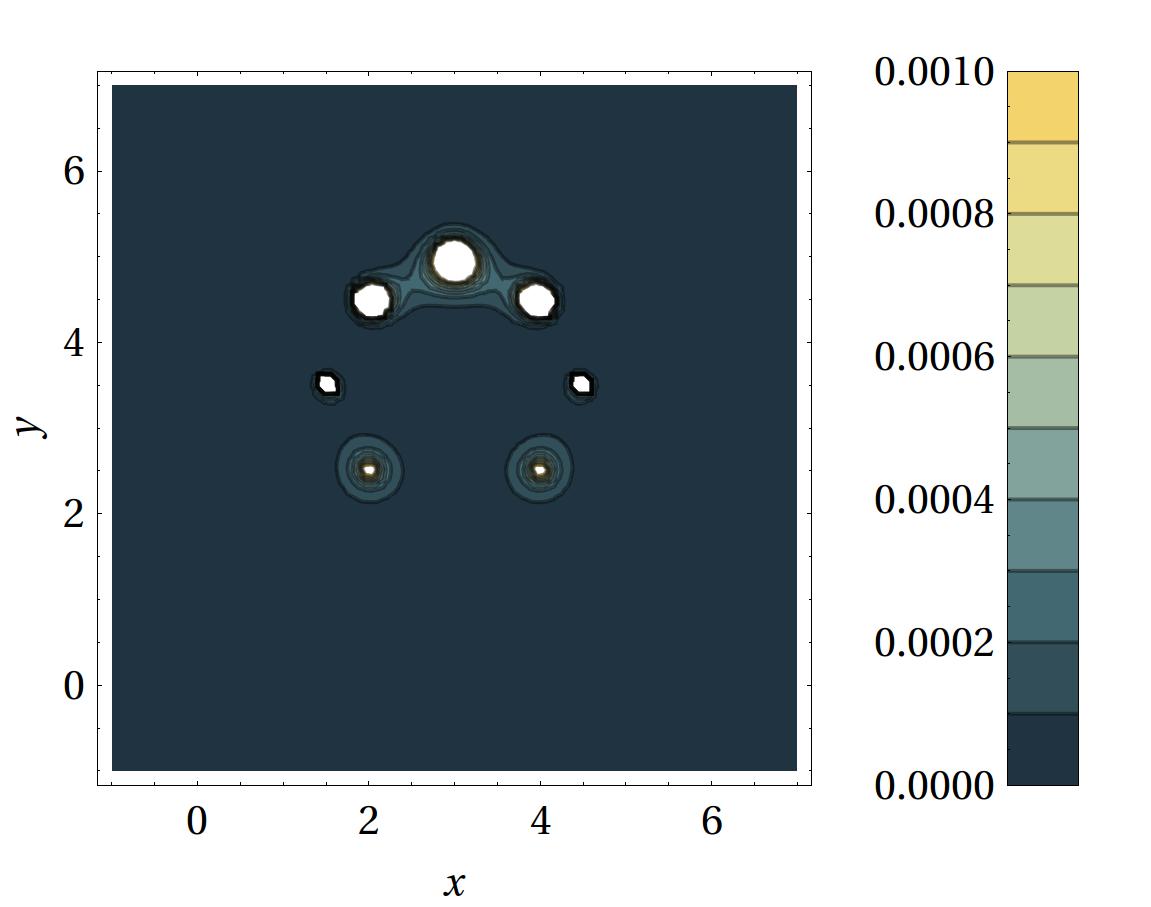} }
	\subfloat[$\mu=1$, $E_y=0.76$]{\label{fig:e} 
	\includegraphics[width=0.3\textwidth]{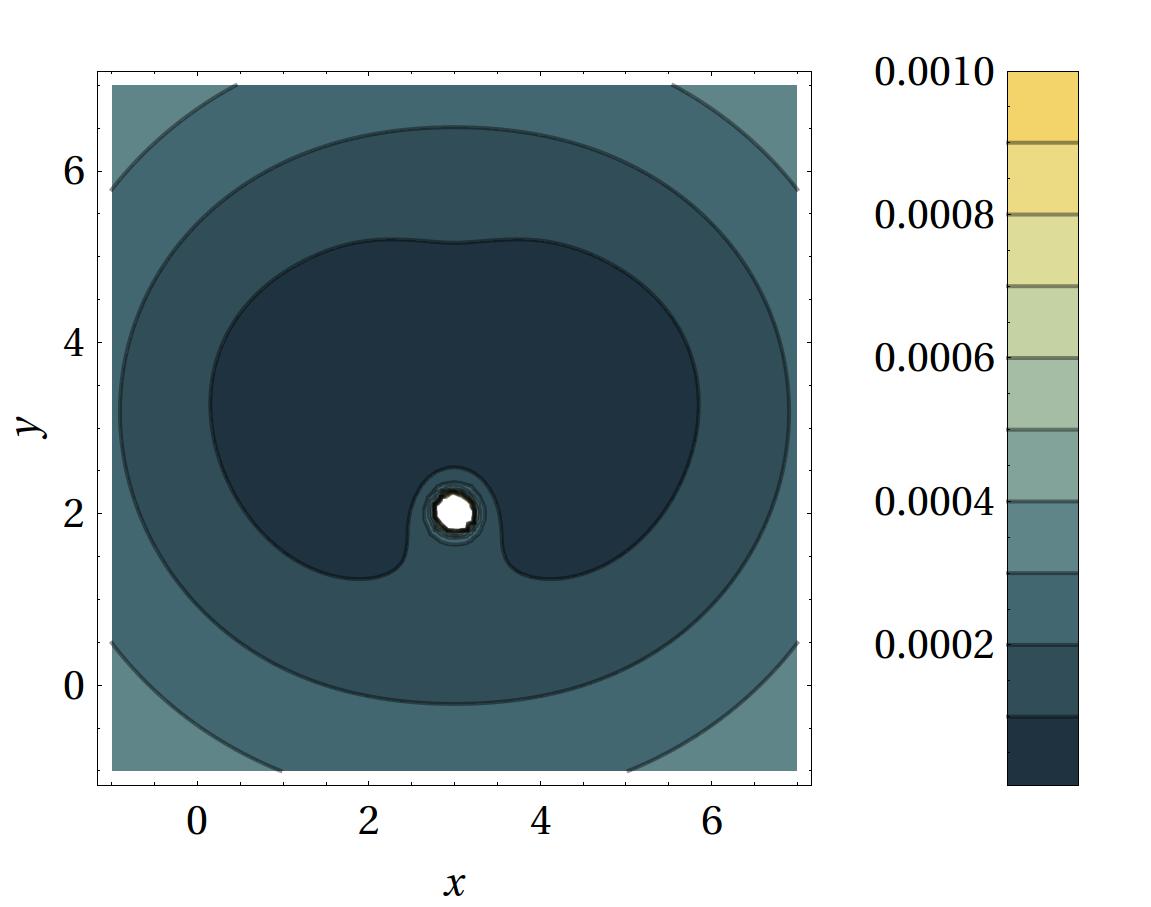} }
	\subfloat[$\mu=1$, $E_y=1$]{\label{fig:f} 
	\includegraphics[width=0.3\textwidth]{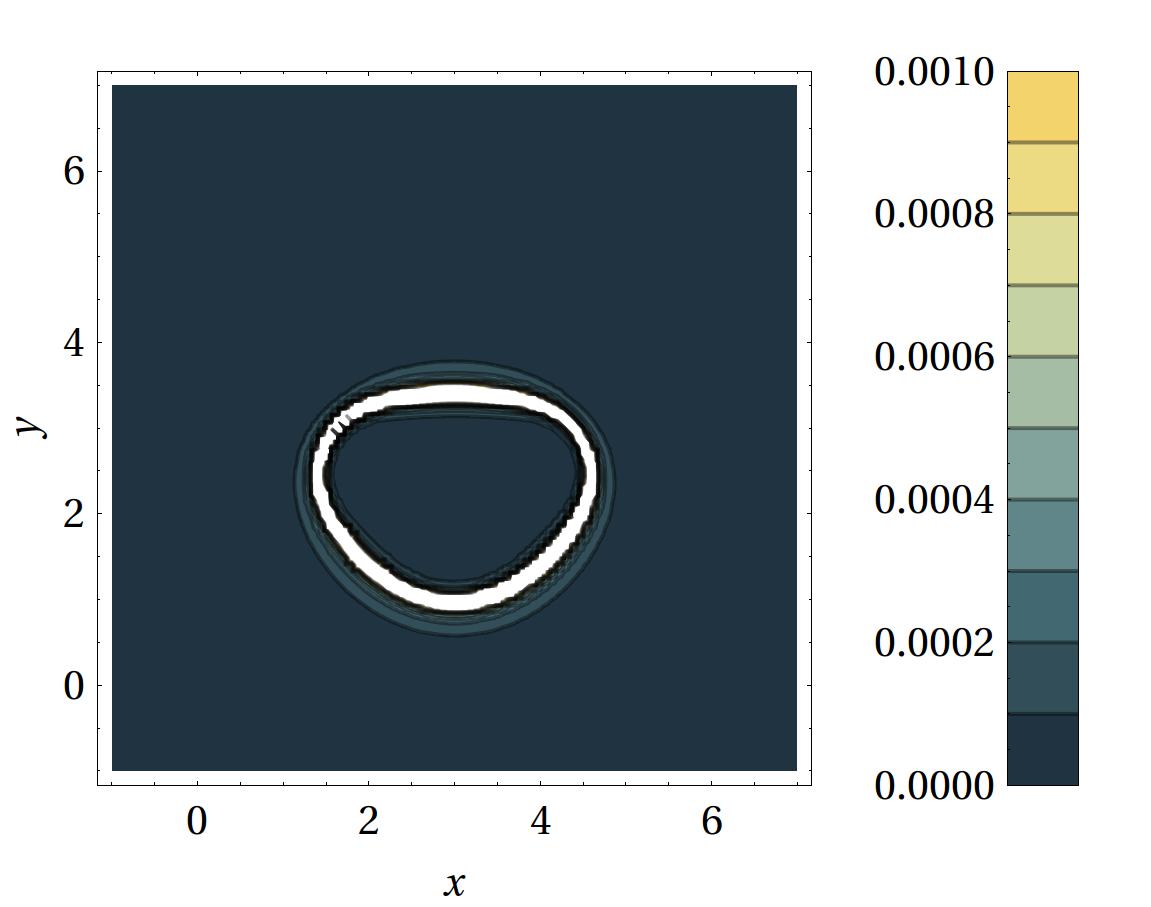} }\\
	\subfloat[$\mu=0.30$, $t_1=0$]{\label{fig:g} 
	\includegraphics[width=0.3\textwidth]{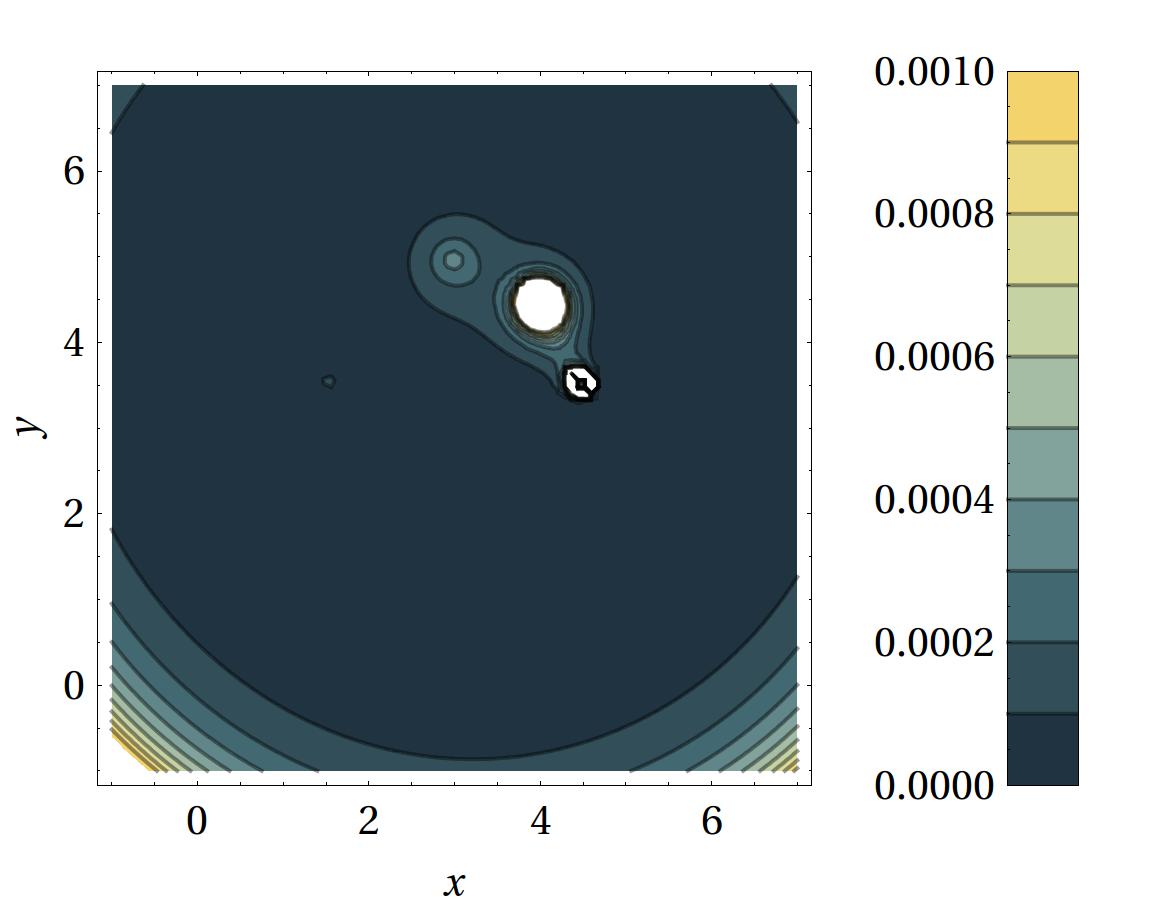} }
	\subfloat[$\mu=-0.40$, $t_1=0$]{\label{fig:h} 
	\includegraphics[width=0.3\textwidth]{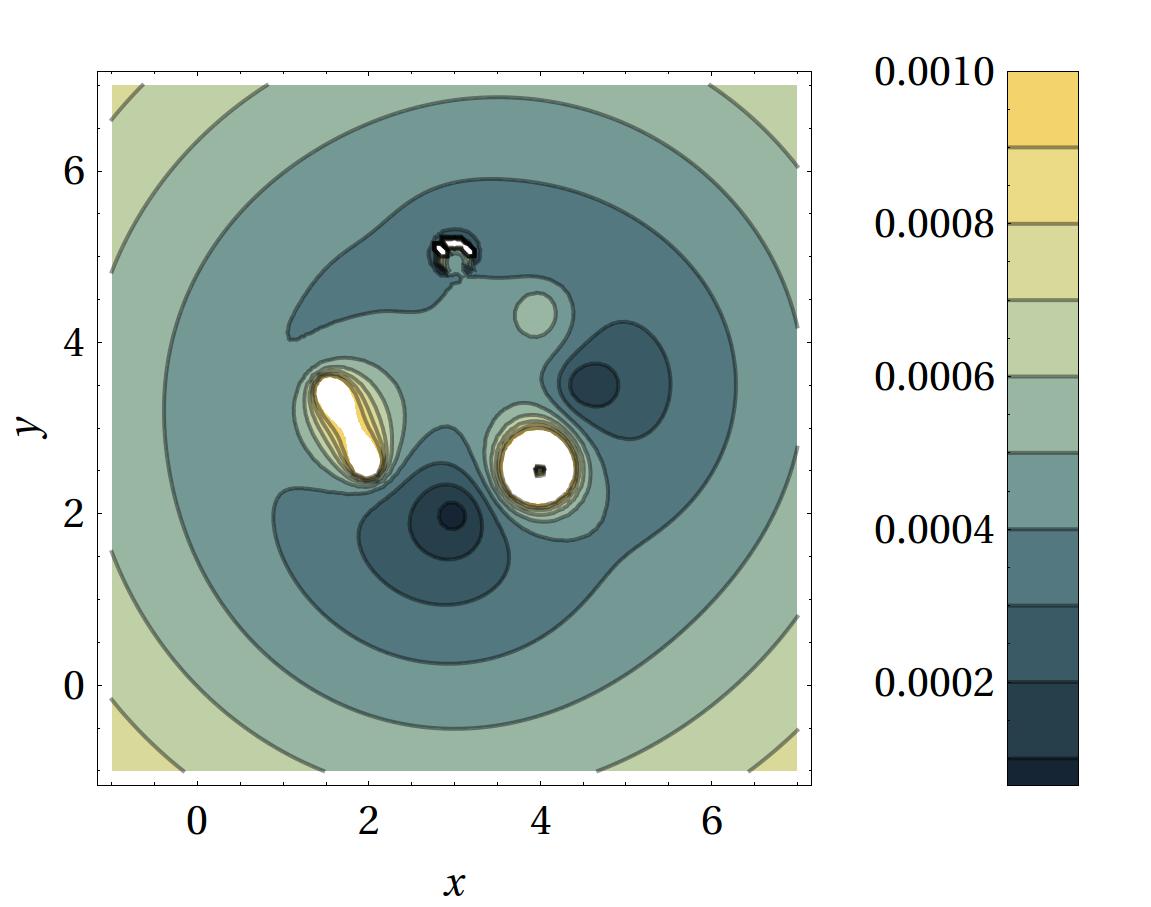} }
	\subfloat[$\mu=-1$, $t_1=0$]{\label{fig:i} 
	\includegraphics[width=0.3\textwidth]{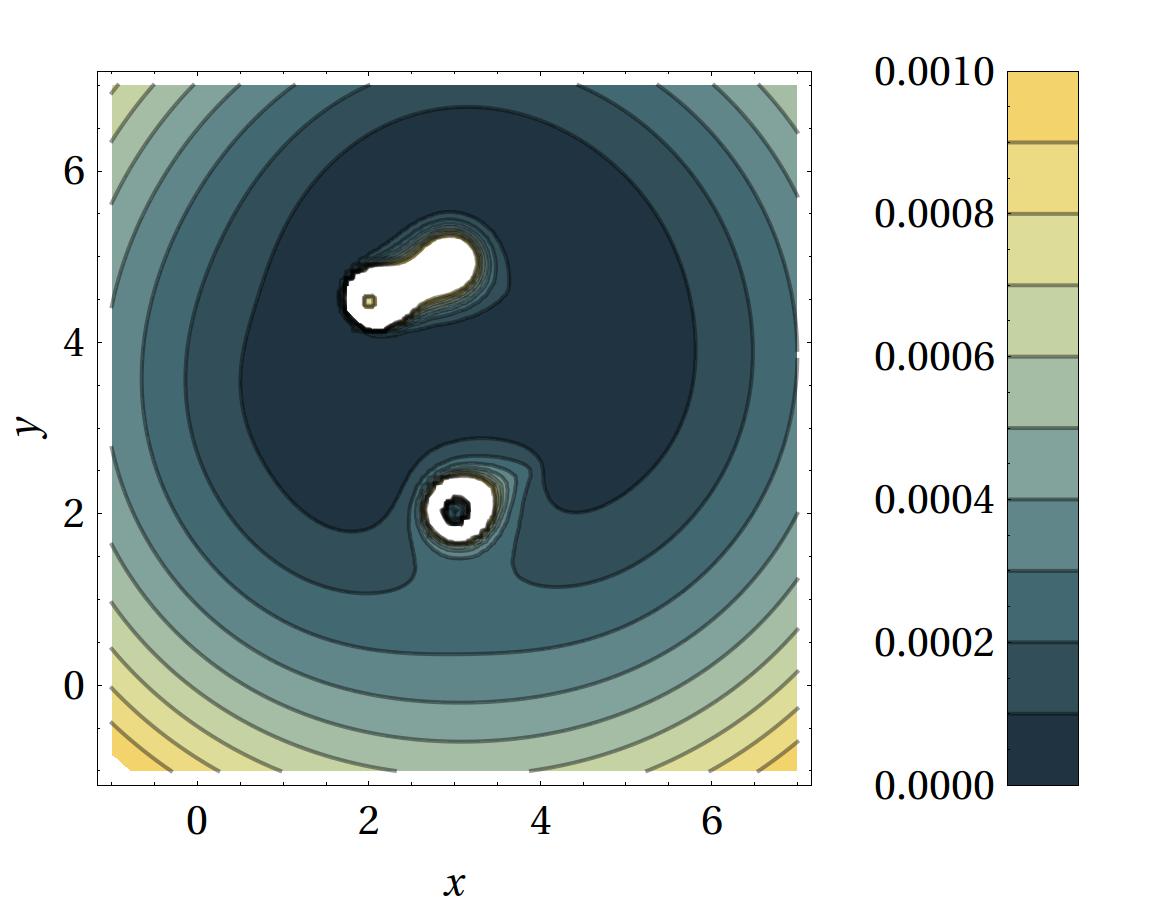} }
	\caption{Conductance contour plots of 
		a ring with 8 sites as a function of the position $(x,y,z)$ of the charged impurity (where $z=0.15$)
		for several values of the chemical 
potential $\mu$: (a), (b) and (c)  in the case of the perfect ring; (d), (e) and (f)  in the case of applied electric field and (g) (h) and (i)  in the case of the broken ring. The color bars indicate the plot range in the contour plots in units of $\frac{e^2}{h}$  and white regions reflect high  conductance values above the upper limit of the plot range.  }
		\label{fig:contourplots}
\end{figure*}

\section{Results}
\label{section:results}
We have applied the method described in the previous section to the determination of the differential conductance  of an open tight-binding ring under electric field and threaded by magnetic flux. We also look into the effect of reducing one of the hoppings integrals in the ring, effectively changing the boundary conditions from periodic to open. 
We also consider the influence of the radial electric field created by a charged impurity  in the neighbourhood of the ring and the results obtained lead us to suggest that the conductance through clusters may be used as a new tool for microscopy, that is, its sensitivity to local electric fields can, in principle, be used to obtain images in a similar way as a 
scanning tunnelling microscope.

In order to better understand the results for the conductance, the energy spectra of the isolated ring is determined.
In the center plot of Fig.~\ref{fig:spectra8ring}, we show the energy spectra of a ring with 8 sites  as a function of the magnetic flux with $\epsilon_i=0$ and nearest neighbor hopping $t=1$, see Fig.~\ref{fig:8ringperfect}. This energy spectra follows exactly the behavior described by Eq.~\ref{AB6}.  The left plot display the same spectra when a planar uniform electric field is present, $eE_y=0.5$, see Fig.~\ref{fig:8ringelectric}. This planar electric field  shifts the on-site energy of every site, ${\epsilon}_j \rightarrow {\epsilon}_j  + e \vec{r}_j \cdot \vec{E} $.
The right plot corresponds to the change of boundary conditions from periodic to open, see  Fig.~\ref{fig:8ringincompleto}. In the left and center plots, the Aharonov-Bohm oscillations of the ground state energy are observed while in the right plot they are absent.
The application of the electric field  leads to two changes in the energy spectra:
(i) The perturbation caused by the electric field has a first order effect 
for degenerated states and a second order effect for non-degenerate levels.
This means that a perturbation produces a "repulsion" between levels that is 
stronger the closer the levels are.
(ii) The influence of the magnetic field decreases as the electric field 
intensity increase. For high enough values of electric field the energy 
oscillations due to the magnetic flux are no longer present and the energy 
dependence on flux becomes flat, reflecting the localization of the particles 
due to the large differences between on-site energies. 

The respective plots of the conductance, evaluated at two values of chemical potential, $\mu=0.01$ and 1 (which are indicated by red and blue dashed lines  in the top plots) are shown in the bottom plots of Fig.~\ref{fig:spectra8ring}.
In the center plot, the Aharonov-Bohm effect in the conductance with a periodicity of one quantum flux is observed as well as zero conductance for $\Phi/\Phi_0=n+1/2$, with integer $n$ (reflecting the destructive interference of the plane waves travelling through the two branches of the ring). The Aharonov-Bohm effect disappears in the broken ring.
The conductance has  peaks when the chemical potential 
has the same value as any of the system eigenvalues. This is called resonant tunnelling and for weak coupling to the leads, these peaks  have  the Breit-Wigner shape \cite{Stone1985,Breit1936}. Note that the conductance profiles are very sensitive to changes in the chemical potential. 

Now, we discuss the influence of non-uniform local electric fields in the conductance through the ring.
In the following, we consider  a  radial electric field  generated by a charged impurity in the proximity of the ring,  according to the Coulomb's law, but other forms of  local electric fields should lead to similar results. We  neglect the possibility of particle hoppings between the cluster and the impurity, as well as the effect of the electric field in the leads. 
As in the case of the uniform electric field, this field changes the on-site 
energy, which is now given by 
${\epsilon}_j \rightarrow \epsilon_j  + \frac{C}{\vert \vec{r_j}-\vec{e}\vert}$,
where $C$ is a constant. The impurity is kept at a fixed distance $z$  of the ring plane (this distance  acts as a cutoff to the Coulomb potential) and its position is swept in the $x$ and $y$ direction, the conductance being calculated for each impurity position. The respective contour plots are shown in Fig.~\ref{fig:contourplots}, for the three cases described by Figs.~\ref{fig:8ringelectric}, \ref{fig:8ringperfect},
and \ref{fig:8ringincompleto}, namely, Figs.~\ref{fig:a}, \ref{fig:b} and \ref{fig:c}  in the case of the perfect ring, Figs.~\ref{fig:d}, \ref{fig:e} and \ref{fig:f}  in the case of the  ring with an applied electric field and Figs.~\ref{fig:g}, \ref{fig:h} and \ref{fig:i}  in the case of the broken ring. In general, we observe in these contour plots,  peaks or dips in the conductance when the impurity $(x,y)$ position coincides with a site of the ring.
In the case of Figs.~\ref{fig:a}, \ref{fig:b} and \ref{fig:c}  (perfect ring), depending on the value of the chemical potential, peaks or dips appear with a symmetric configuration, which reflects two factors: (i) the  probability  density of the eigenstates of the ring Hamiltonian and in particular of the eigenstate with energy nearest to the chemical potential; (ii) the influence of the leads which effectively changes the on-site energy of the two contact sites of the ring. Qualitatively, one may say that  the leads lift the degeneracy of the energy spectra of the ring and since this effect is local, the standing waves in the ring with the same momentum acquire different energies (the difference is small for weak coupling to the leads). Therefore, one can infer from the conductance contour plots in the case of the symmetric ring, the probability density of these standing waves.

When an electric field is applied, the eigenstates of the ring have  less symmetric probability density  profiles,  
with more probability density being accumulated in the ring in the direction of the electric field or in the opposite direction, depending on the energy of the eigenstate. This is reflected by the conductance contour plots of Figs.~\ref{fig:d}, \ref{fig:e} and \ref{fig:f}, where the lowering of symmetry is observed. Note that for fixed chemical potential, varying the electric field, the  energy of the eigenstates of the ring  is increased and several crossings of the eigenvalues of the ring with chemical potential occur as shown in Fig.~\ref{fig:spectra8ring}, so that in the conductance contour plots,  probability density distributions corresponding to different eigenstates of the ring are observed.
Breaking a hopping link in the tight-binding ring has even a stronger effect in the eigenstates probability densities of the ring. In Figs.~\ref{fig:g}, \ref{fig:h} and \ref{fig:i}, the absence of symmetry is clearly observed. Again, adjusting the chemical potential, the conductance of the broken ring become more sensitive to local electric fields in different regions of the $xy$ plane.

The  sensitivity to local electric fields as well as the possibility of probing different points of space adjusting the chemical potential, lead us to suggest that the conductance through clusters may be used as a new tool for microscopy, that is, it can, in principle, be used to obtain images in a similar way as a 
scanning tunnelling microscope.
Note that small changes in the vertical position of 
the impurity do not change significantly the conductance shape. The farther 
the impurity is from the cluster plane, the less it couples to the cluster and the smoother the conductance plot gets.

\section{Conclusion}
\label{section:conclusion}

In this paper, we showed how to determine numerically the  effects of applied magnetic and electric fields on the conductance of tight-binding  quantum clusters connected to one-dimensional leads, using a  simple tight-binding method which could be programmed by an undergraduate student attending a Solid State course or  a final year project.
Important phenomena of the  transport through nanorings such as Aharonov-Bohm conductance oscillations, resonant tunnelling or destructive interference are observed as magnetic field, chemical potential or other parameters are varied.
We  explored in particular the effect of a charged impurity on the conductance through the ring.
We  observed that the conductance has a strong sensitivity to the local electric 
field and  suggested that it can, in principle, be used to obtain images in a similar way as the 
scanning tunnelling microscope.

AAL was  supported by 
Funda\c c\~ao para a Ci\^encia e a Tecnologia (Portugal), co-financed by FSE/POPH, under grant SFRH/BD/68867/2010 and of the Excellence Initiative of the German Federal and State Governments (grant ZUK 43)
\\
\\

\bibliography{conductance}

\end{document}